\documentclass[conference]{IEEEtran}
\IEEEoverridecommandlockouts
\usepackage{cite}
\usepackage{amsmath,amssymb,amsfonts}
\usepackage{algorithmic}
\usepackage{graphicx}
\usepackage{textcomp}
\usepackage{xcolor}
\usepackage[misc]{ifsym}

\def\BibTeX{{\rm B\kern-.05em{\sc i\kern-.025em b}\kern-.08em
    T\kern-.1667em\lower.7ex\hbox{E}\kern-.125emX}}
\begin{document}
% CSVAR: Channel-Wise Image Shuffling with Variance-Guided Adaptive Region Partitioning for 
\title{CSVAR: Enhancing Visual Privacy in Federated Learning via Adaptive Shuffling Against Overfitting}

\author{Zhuo Chen, 
Zhenya Ma, 
Yan Zhang, 
Donghua Cai, 
Ye Zhang, 
Qiushi Li, 
% Qiushi Li*, 
Yongheng Deng, 
Ye Guo,\\
Ju Ren and 
Xuemin (Sherman) Shen
}

\maketitle

\def\thefootnote{}\footnotetext{Both authors (Zhuo Chen and Zhenya Ma) contributed equally to this work.}
\def\thefootnote{}\footnotetext{Zhuo Chen and Ye Guo are with China Mobile Communications Group Co.,Ltd. (email: chenzhuoit@chinamobile.com, guoye@chinamobile.com)}
\def\thefootnote{}\footnotetext{Zhenya Ma, Yan Zhang, Donghua Cai, Qiushi Li (corresponding author) and Yongheng Deng are with the Department of Computer Science and Technology, Tsinghua University, China. (email: 
mzy23@mails.tsinghua.edu.cn, 
yan-zhan23@mails.tsinghua.edu.cn, 
caidonghua2003@gmail.com, 
lqs@tsinghua.edu.cn, 
dyh2024@tsinghua.edu.cn)
}
\def\thefootnote{}\footnotetext{Ye Zhang is with the School of Computer Science, Beijing Information Science and Technology University, China. (email: yezhang@bistu.edu.cn)
}
\def\thefootnote{}\footnotetext{Ju Ren is with the Department of Computer Science and Technology, Tsinghua University, China, and Zhongguancun Laboratory, Beijing, China. (email: renju@tsinghua.edu.cn)
}
\def\thefootnote{}\footnotetext{Xuemin (Sherman) Shen is with the Department of Electrical and Computer Engineering, University of Waterloo, Canada. (email: sshen@uwaterloo.ca)
}
% \def\thefootnote{\textsuperscript{\Letter}}\footnotetext{Corresponding author (email: lqs@tsinghua.edu.cn).}

% \begin{abstract}
% This document is a model and instructions for \LaTeX.
% This and the IEEEtran.cls file define the components of your paper [title, text, heads, etc.]. *CRITICAL: Do Not Use Symbols, Special Characters, Footnotes, 
% or Math in Paper Title or Abstract.
% \end{abstract}
\begin{abstract}
Although federated learning preserves training data within local privacy domains, the aggregated model parameters may still reveal private characteristics. This vulnerability stems from clients' limited training data, which predisposes models to overfitting. Such overfitting enables models to memorize distinctive patterns from training samples, thereby amplifying the success probability of privacy attacks like membership inference. 
% While federated learning (FL) preserves data privacy by keeping raw images on client devices, two critical vulnerabilities remain: (1) exposure of private training images during client-side transmission between collection and computation nodes, and (2) privacy leakage through model updates as clients overfit to local data. Existing solutions like differential privacy (DP) cannot effectively protect visual features of exposed images and degrade model utility significantly. 
To enhance visual privacy protection in FL, we present CSVAR(Channel-Wise Spatial Image Shuffling with Variance-Guided Adaptive Region Partitioning), a novel image shuffling framework to generate obfuscated images for secure data transmission and each training epoch, addressing both overfitting-induced privacy leaks and raw image transmission risks. 
CSVAR adopts \textit{region-variance} as the metric to measure visual privacy sensitivity across image regions. Guided by this, CSVAR adaptively partitions each region into multiple blocks, applying fine-grained partitioning to privacy-sensitive regions with high region-variances for enhancing visual privacy protection and coarse-grained partitioning to privacy-insensitive regions for balancing model utility. In each region, CSVAR then shuffles between blocks in both the spatial domains and chromatic channels to hide visual spatial features and disrupt color distribution. Experimental evaluations conducted on diverse real-world datasets demonstrate that CSVAR is capable of generating visually obfuscated images that exhibit high perceptual ambiguity to human eyes, simultaneously mitigating the effectiveness of adversarial data reconstruction attacks and achieving a good trade-off between visual privacy protection and model utility. %Extensive experiments on medical imaging and facial recognition benchmarks demonstrate CSVAR’s superior privacy-utility tradeoffs—reducing reconstruction accuracy by 62\% compared to DP-based methods while maintaining within 2\% of baseline model performance. The framework’s client-side processing also prevents raw data exposure during internal transmissions, offering comprehensive protection throughout the FL pipeline.

\end{abstract}

\section{Introduction}
Federated learning (FL\cite{yue2024federated}) has emerged as a promising decentralized learning paradigm that enables multiple participants to collaboratively train a shared model without sharing their raw data. Instead of centralizing sensitive data in the cloud, FL allows clients to train locally and only exchange model updates (e.g., gradients or weights) with a central server for aggregation. This framework enhances privacy by preserving training data within local privacy domains, while maintaining model performance. Thus, FL is particularly valuable for vision-based applications like medical imaging analysis, facial recognition systems, and mobile photography services where protecting visual privacy and preventing reconstruction of identifiable images or extraction of sensitive visual features are paramount.

Despite its privacy-aware design, FL still faces two critical vulnerabilities for visual data protection. First, although clients only share model updates rather than raw images with the remote server, the typically small and non-IID local datasets \cite{deng2021auction} often lead to severe overfitting during local training. This causes model weights to encode excessive information about specific training samples, enabling adversaries to visually reconstruct private images through Model Inversion Attacks \cite{DBLP:conf/cvpr/ZhangJP0LS20} (e.g., recovering patient faces from medical image models) or determine whether a given image belongs to training sets through Membership Inference Attacks \cite{DBLP:conf/sp/ShokriSSS17}. Second, in vision-centric deployments (e.g., healthcare systems with CT/MRI scanners, smart cameras), a security gap exists between image collection sensors and computing nodes. The transmission of private raw images within local networks creates attack surfaces where images could be intercepted, undermining FL's end-to-end privacy guarantees for private data \cite{geiping2020inverting}.

Differential privacy (DP) \cite{DBLP:conf/ccs/AbadiCGMMT016} has been widely adopted in FL to mitigate such privacy risks through introducing carefully calibrated random noise into training processes. For visual data protection, this randomness ensures each training iteration operates on effectively varied versions of the input images, alleviating the overfitting phenomenon. However, DP's noise-based protection operates in the high-frequency domain - while mathematically sound for membership privacy, the human eyes can easily filter such perturbations, leaving sensitive image features exposed \cite{DBLP:conf/ndss/LiZ00Z24}. What's worse, achieving strong visual privacy through DP often requires significant noise injection, which often degrades model utility unacceptably.

Building upon these limitations of DP, we explore the image shuffling mechanism that can generate obfuscated images for data transmission and each training epoch, addressing both overfitting-induced privacy leaks and raw image transmission risks. However, designing an effective shuffling strategy faces several nuanced challenges. 
First, there exists the non-uniform nature of visual privacy across different image regions: sensitive regions like faces or medical identifiers require stronger obfuscation than structural features like backgrounds. This necessitates an adaptive shuffling granularity of a given region: overly fine-grained shuffling, like pixel-level shuffling, maximally alleviates overfitting but destroys essential spatial structures needed for model learning, whereas overly coarse shuffling risks insufficient privacy protection. We need to divide the region into smaller blocks, shuffle between them, and preserve the pixels within the block. A one-granularity-fits-all shuffling approach would either over-protect unimportant areas or under-protect sensitive ones, motivating the need for spatially adaptive transformations to make the trade-off.
Second, we need a principled metric to measure a region's visual privacy sensitivity (human-recognizable features) and model overfitting risk. Without such guidance, naive shuffling implementations either compromise privacy through insufficient obfuscation or degrade model utility unacceptably via excessive shuffling. 
Third, while conventional pixel-wise spatial shuffling provides basic spatial obfuscation, it fails to disrupt color distributions - a critical vulnerability as attackers can reconstruct sensitive features through chromatic analysis.

To address the challenges outlined above, % we propose CSVAR (Channel-Wise Spatial Image Shuffling with Variance-Guided Adaptive Region Partitioning), a novel shuffling mechanism for anti-overfitting that dynamically adjusts shuffling granularity based on different regions' variances, which can prevent overfitting-induced visual privacy leakage and the client-side data transmission under FL, meanwhile. 
we propose CSVAR (Channel-Wise Spatial Image Shuffling with Variance-Guided Adaptive Region Partitioning), a novel anti-overfitting mechanism that employs dynamic region-adaptive shuffling. By preventing model overfitting through variance-guided image shuffling, CSVAR simultaneously prevents overfitting-induced visual privacy leakage and client-side data transmission under FL.CSVAR will first divide the images into regions and calculate each region's variance. Guided by variances, CSVAR will partition the region into smaller blocks adaptively and shuffle between blocks both in the spatial domain to destroy spatial visual features and the chromatic channel space to disrupt color distributions. Specifically, CSVAR will partition a privacy-sensitive region with high variance into small blocks to enhance privacy protection. Meanwhile, for a privacy-insensitive region with low variance, CSVAR will partition it with much bigger blocks to preserve model utility. CSVAR ensures that: 1) private training images are protected during client-side transmission by transmitting obfuscated versions instead of raw images, 2) different shuffled versions are used in different epochs to alleviate overfitting and overfitting-induced privacy leakage, and 3) the adaptive shuffling approach maintains a careful balance between visual privacy protection and model utility throughout the federated learning process.

% To evaluate ... Results show that...

Our contributions are summarized as follows:
\begin{itemize}
    \item We propose CSVAR, a novel image shuffling framework to prevent visual privacy leakage from overfitting and client-side data transmission under Federated Learning.
    \item We adopt region-variance as the metric to quantify the visual privacy sensitivity of different regions of an image and guide the shuffling granularity.
    \item We adaptively partition each image region into smaller blocks with different granularity guided by region-variance, and then shuffle between blocks in both the spatial domains and chromatic channels.
    \item We conduct extensive experiments on real-world datasets, demonstrating that CSVAR can generate visually obfuscated images that exhibit high perceptual ambiguity to human eyes, mitigate the effectiveness of adversarial data reconstruction attacks, and achieve a good trade-off between visual privacy protection and model utility.
\end{itemize}

\section{Threat Model and  Design Goals}
This section formalizes our system model and threat model, followed by our design goals.
\subsection{System Model}
We consider a conventional federated learning framework comprising a central server and multiple distributed clients, composed of two functionally distinct components:  local data collection and computation nodes. This architecture mirrors a lot of real-world deployments, particularly in privacy-sensitive domains like healthcare, where medical imaging scanners (CT/MRI, lacking computational capabilities) must send collected data to local computation nodes through internal networks to perform local model training.

During each training epoch, the system operates through several key phases. Initially, data collection nodes gather raw local training data through their sensing capabilities, then transfer private data to co-located computation nodes via internal networks. The computation nodes subsequently perform local model training using their private datasets before submitting parameter updates to the central server. Following aggregation of clients' updates, the server distributes the updated global model back to all participating clients for the next training epoch. 

\subsection{Security Model}
We focus on two key attack scenarios for image data protection in the above system model. First, attackers can intercept client-side data transmissions between data collection nodes and computation nodes during internal network transfers. Second, attackers can access the server's global model weights. Here, the attacker can be the curious-but-honest servers that honestly perform federated learning tasks but attempt to extract private data from model weights by employing techniques such as GAN-based data reconstruction attacks to reconstruct private training images or membership inference attacks to identify the ownership of the data used in the training process. These scenarios cover both raw data exposure during local transfers on the client side and potential privacy leaks through model weights on the server side.
\subsection{Design Goals}
Our framework aims to achieve three fundamental objectives that alleviate overfitting, and preserve the visual privacy of image data while retaining the utility of the model:

\textbf{Secure data transmission between data collection and computation nodes on the client side.}  Our framework should provide robust obfuscation to the private image data to avoid privacy leakage even if the attacker can access obfuscated data.

\textbf{Defend against visual privacy leakage from the overfit model on the server side.} Our proposed method should address the overfitting-induced privacy leakage on the private client data: (1)preventing data reconstruction attacks from reconstructing private images and (2) resisting membership inference attacks that attempt to identify training data participation.

\textbf{Achieve satisfactory trade-off between visual privacy protection and model utility. } Our framework should aggressively protect privacy-sensitive areas while maintaining structural integrity in privacy-insensitive regions, to minimize accuracy degradation while meeting privacy requirements.

\section{Motivation}
\subsection{Overfitting-Induced Privacy Leakage in FL}
\begin{figure}
    \centering %图片居中
    \includegraphics[width=0.4\textwidth]{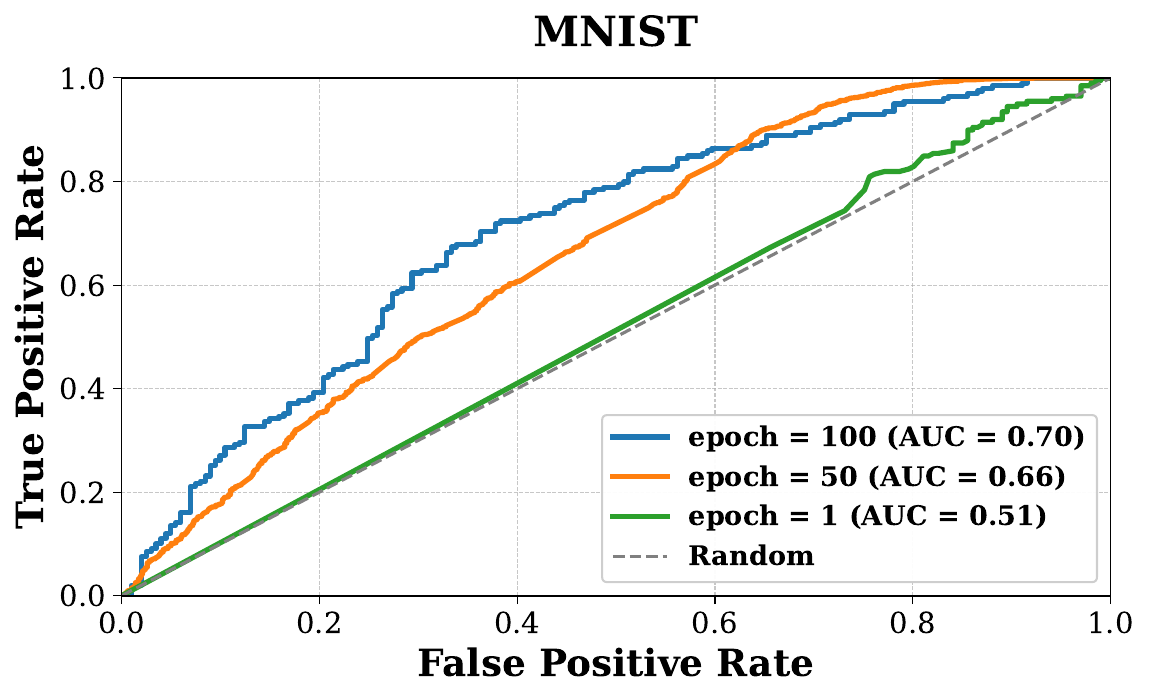}
    %插入图片，[]中设置图片大小，{}中是图片文件名
    \vspace{-0.1in}
    \caption{ROC results of Membership Inference Attacks Varying Degrees of Overfitting. Low AUC means low attack Success Rate.} %最终文档中希望显示的图片标题
    \label{img:mia_overfitting} %用于文内引用的标签
    \vspace{-0.1in}
\end{figure}
Federated learning systems are particularly vulnerable to overfitting due to the typically small and non-IID nature of local client datasets. 
As training progresses, this overfitting causes model weights to encode increasingly sensitive features about private training samples, and we verify the growth of privacy leakage through \textbf{membership inference attacks(MIA) \cite{DBLP:conf/sp/ShokriSSS17}}. 
Taking advantage of the model weights after training, MIA aims to determine whether a given data point belongs to the training dataset. Figure \ref{img:mia_overfitting} demonstrates that moderately trained models have low MIA success rates (showing accuracy with low AUC near 0.5 similar to Random Guess), while heavily overfit models show high success rates (high AUC of 0.70).

% Through systematic evaluation of two representative attacks in Figure \ref{img-overfitting}, we reveal this concerning relationship:
% \begin{itemize}
% \item \textbf{GAN-based reconstruction attacks \cite{DBLP:conf/ccs/HitajAP17}} demonstrate progressively clearer feature extraction across training epochs. Early-stage models with minimal overfitting produce only noisy, unrecognizable reconstructions, while late-stage overfit models reveal identifiable facial features from the AT\&T dataset and distinct digit patterns from MNIST. This progression directly correlates with the growing disparity between training and validation accuracy.
% \item The growth of privacy leakage is further evidenced through \textbf{membership inference attacks \cite{DBLP:conf/sp/ShokriSSS17}}. While moderately trained models resist membership inference attempts (showing accuracy near random chance), heavily overfit models permit near-perfect determination of training set membership.
% \end{itemize}
% The findings demonstrate a pronounced positive correlation between the extent of model overfitting and the corresponding attack success rates. 
This motivates our key insight: by shuffling training images for each epoch, we can ensure models encounter varying versions of the data, thereby simultaneously mitigating overfitting and reducing associated privacy leakage risks.

\subsection{Region-Variance as Visual Privacy Indicator}

Visual privacy protection requires recognizing that privacy exhibits an inherent spatial non-uniform nature: while privacy-sensitive regions(e.g., facial features) demand strong obfuscation, homogeneous backgrounds can tolerate lighter protection. An intuition that the variance of a given region can be used to measure privacy-sensitivity, based on the observation that sensitive areas typically exhibit higher pixel-value variance due to complex textures and edges, whereas uniform backgrounds show minimal variance.

\begin{figure}
    \centering %图片居中
    \includegraphics[width=0.48\textwidth]{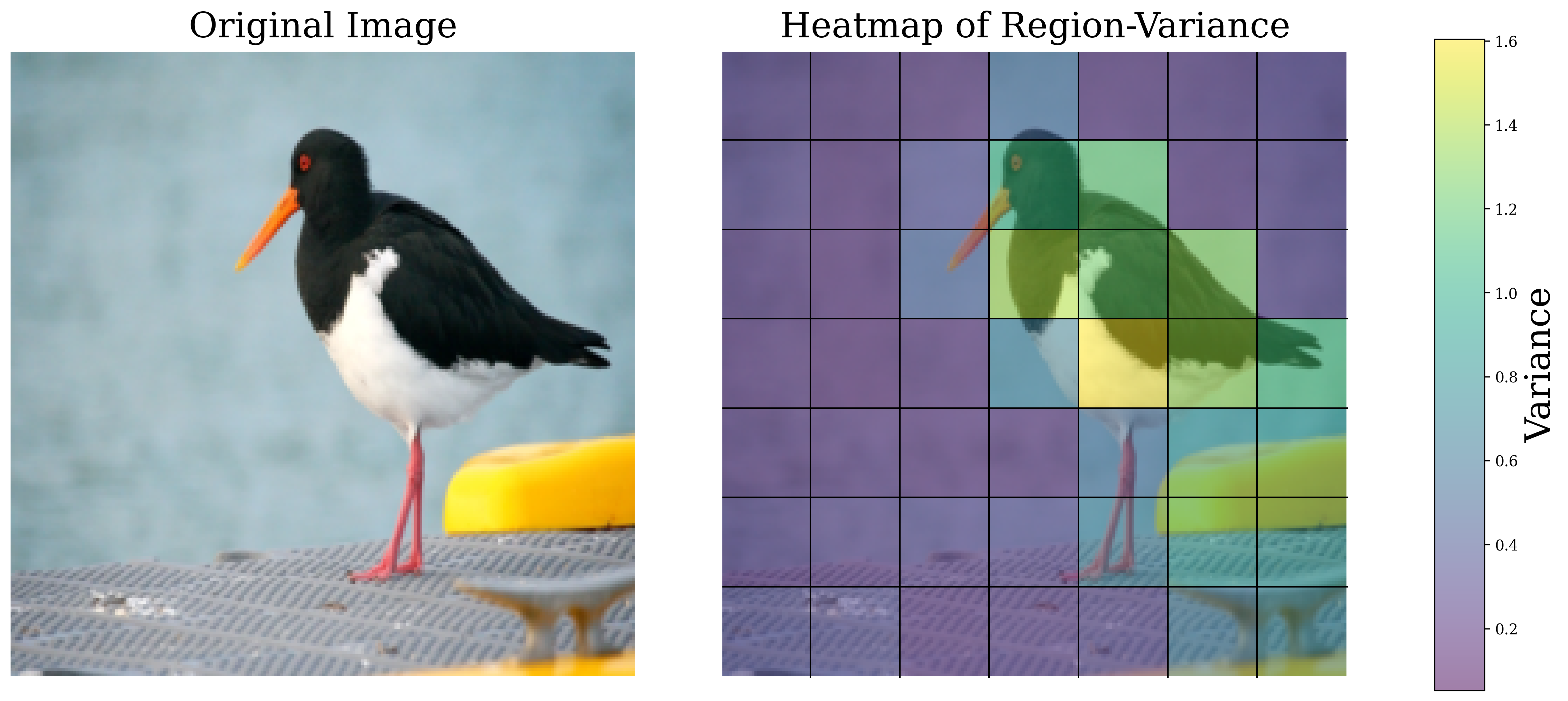}
    %插入图片，[]中设置图片大小，{}中是图片文件名
    \vspace{-0.1in}
    \caption{Region-Variance of different regions in the image of a bird. Lighter region means a higher region-variance.} %最终文档中希望显示的图片标题
    \label{img-var-of-region} %用于文内引用的标签
    \vspace{-0.15in}
\end{figure}

To verify this intuition, our variance computation follows three steps: (1) partitioning the image into 14×14 regions (16×16 pixels each), (2) calculating per-channel variance across all pixels within each region, and (3) averaging variances from each channel. As shown in Figure \ref{img-var-of-region}, results show that high-variance regions (lighter areas in the heatmap) consistently correspond to semantically sensitive features (like the bird's beak/body in Figure \ref{img-var-of-region}). The strong alignment between variance and visual sensitivity confirms region-variance's suitability for guiding protection strategies, thus, we propose region-variance as a quantifiable metric for privacy sensitivity.

\subsection{Shuffling Granularity Trade-off}

\begin{figure}
    \centering %图片居中
    \includegraphics[width=0.48\textwidth]{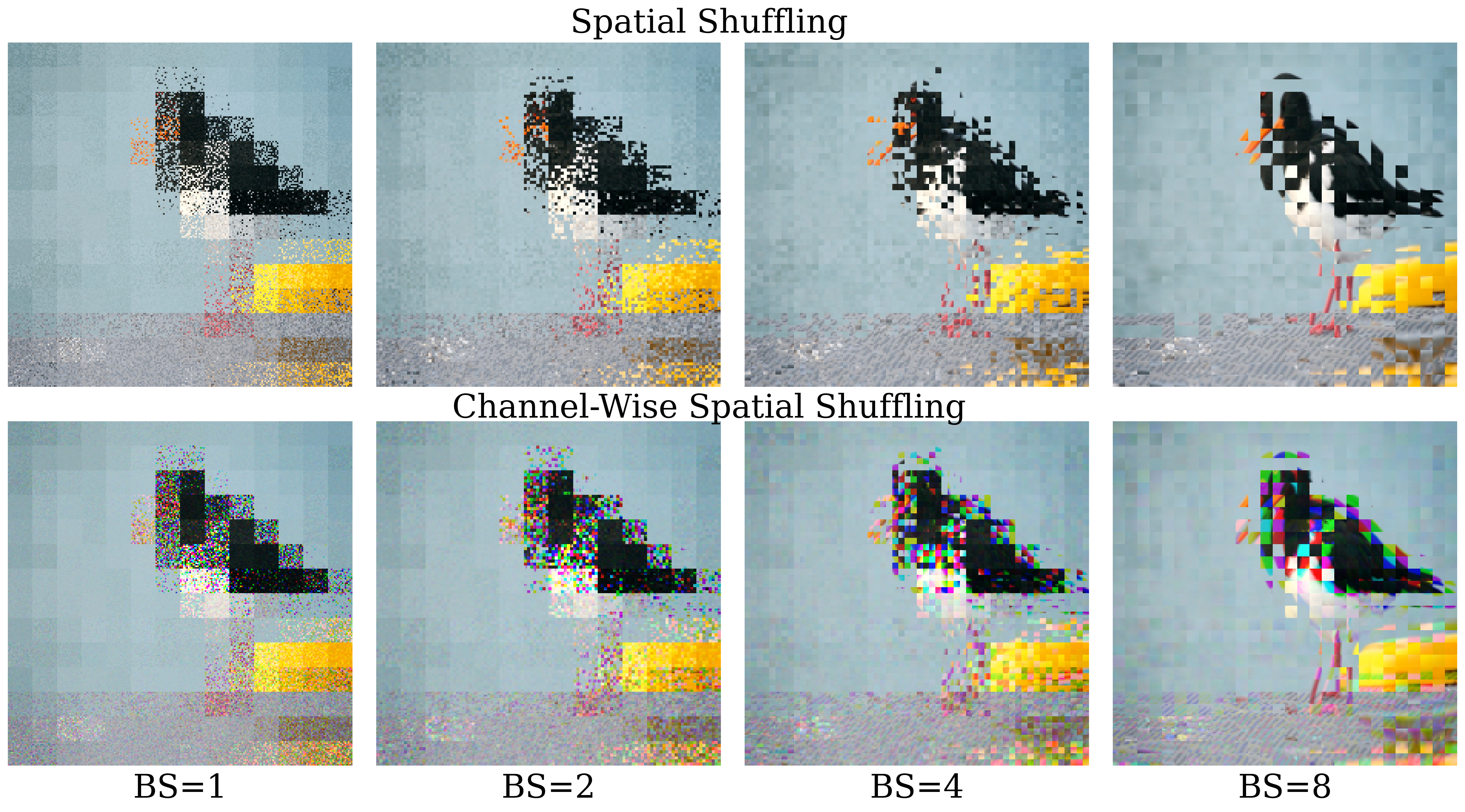}
    %插入图片，[]中设置图片大小，{}中是图片文件名
    \vspace{-0.1in}
    \caption{Visual privacy protection effect under different block size(BS) with spatial-only shuffling and channel-wise spatial shuffling. Results show we can recognize fewer visual features when using a small BS with channel-wise spatial shuffling than a large BS with spatial-only shuffling.} %最终文档中希望显示的图片标题
    \label{img-window-size} %用于文内引用的标签
    \vspace{-0.2in}
\end{figure}

\begin{figure*}
    \centering %图片居中
    \includegraphics[width=0.9\textwidth]{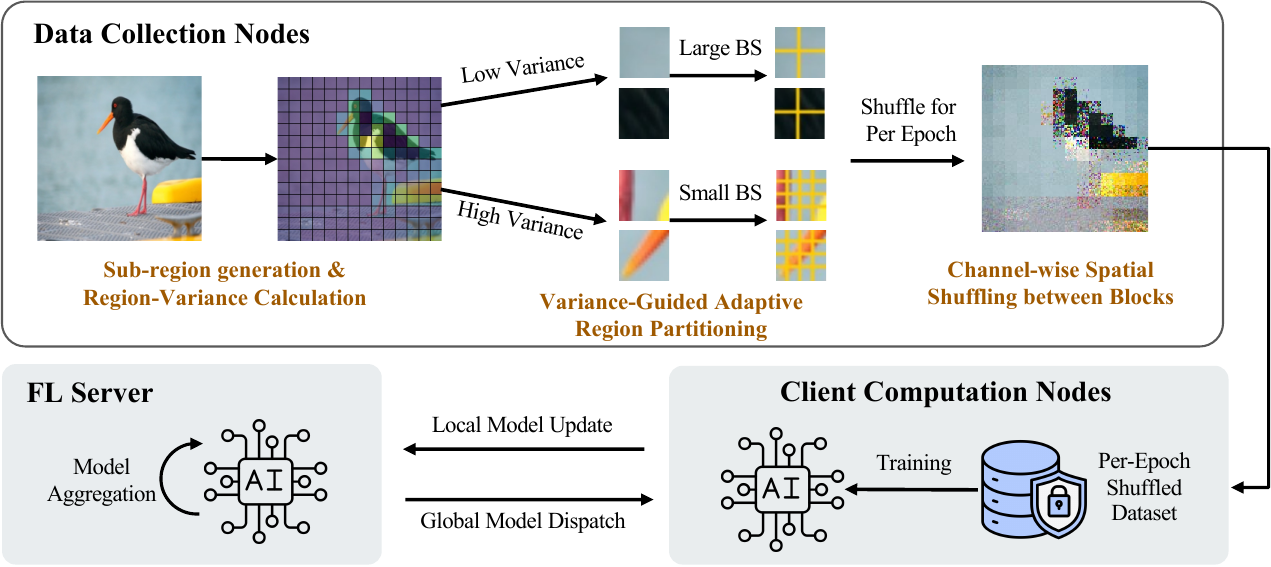}
    %插入图片，[]中设置图片大小，{}中是图片文件名
    \vspace{-0.1in}
    \caption{The System Overview of CSVAR. Note that light color in the heatmap denotes a high variance region. BS means block size.} %最终文档中希望显示的图片标题
    \label{img-system} %用于文内引用的标签
    \vspace{-0.2in}
\end{figure*}

The effectiveness of visual privacy protection through shuffling depends critically on shuffling granularity. As shown in Figure \ref{img-window-size}, we systematically explore this by the following steps: (1) dividing each 224×224 image into 14×14 base regions(16×16 pixels in each region), (2) further splitting each region into BS×BS blocks ($Block Size(BS) \in [1,2,4,8]$), and (3) shuffling between blocks while preserving intra-block pixels. Columns varying with different BS in figure \ref{img-window-size} demonstrate that we can recognize more visual features with a large shuffle granularity(BS=8), which means we can get more visual privacy protection with small BS (e.g., BS=1). This is because pixels within each block will not be shuffled. Small BS provides stronger visual obfuscation by thoroughly disrupting spatial relationships, but at the cost of damaging semantically important structures needed for model learning. Conversely, large blocks maintain better model utility but leak more recognizable features. One key insight we obtain from this is that an effective shuffling strategy requires adaptive BS to partition different regions - using small BS for regions with high region-variances and large BS elsewhere. 

Furthermore, the second row of Figure \ref{img-window-size} shows the shuffled image after applying spatial shuffling to each RGB channel. Results show this channel-wise spatial shuffling provides additional visual privacy protection by disrupting color distributions, thereby preventing potential reconstruction attacks through chromatic analysis.

\section{System Design}
% Building upon the motivation and design goals above, we present the system architecture of CSVAR, our proposed novel framework for enhancing visual privacy in federated learning. The solution addresses privacy leakage risks from both client-side data transmission and server-side privacy leakage from model weights through effective image shuffling.
\subsection{System Overview}
% shuffling before each epoch offline
Building upon the motivation and design goals above, we present the system architecture of CSVAR. Figure \ref{img-system} illustrates the end-to-end workflow of CSVAR. The process begins at the client data collection nodes, where raw private images are applied offline shuffling to generate distinct obfuscated versions for each training epoch and data transmission. First, CSVAR partitions each image into multiple regions and computes their region-variances as the privacy sensitivity metric. Guided by region-variance, CSVAR then adaptively partitions each region: privacy-sensitive regions with high variance will be divided into smaller blocks to maximize obfuscation, while low-variance regions with less sensitive features are split into larger blocks to better preserve model utility. Finally, CSVAR performs block-wise spatial shuffling, which shuffles between blocks both in the spatial domain to destroy spatial visual features and the chromatic channel space to disrupt color distributions. These shuffled images are then transmitted to the client computation nodes for local model training. 

During the online federated learning phase, each training epoch follows a distributed workflow where participating computation nodes execute local model training on epoch-specific shuffled images. These clients subsequently transmit their local model updates to the central server. Upon receiving updates from all clients, the server employs a secure aggregation protocol to compute the global model update, which is then redistributed to participating nodes for the subsequent epoch. Notably, each training epoch incorporates distinct versions of shuffled image data to alleviate model overfitting.

This design ensures that: (1) Private images are protected during client-side transmission by sending obfuscated versions instead of raw images, (2) different shuffled versions are used in different epochs to alleviate overfitting and overfitting-induced privacy leakage, and (3) the adaptive approach maintains a careful balance between visual privacy protection and model utility throughout the federated learning process.

\subsection{Design Details}
\textbf{1) Sub-region Generation \& Region-Variance Calculation.}
CSVAR begins by partitioning each input image $I \in \mathbb{R}^{H\times W\times C}$ into non-overlapping regions $R_{i,j}$ of size $S \times S$ pixels, where $S$ is determined adaptively based on image Height(H) and Width(W):

\begin{equation}
S = 2^{\lceil \log_2(\sqrt{\max(H,W)}) \rceil}
\label{eq:region_size}
\end{equation}

This choice of region size ensures: (1) sufficient granularity for privacy protection while maintaining recognizable local features ($S=16$ for standard $224\times224$ images), (2) power-of-two region sizes enable natural binary partitioning(like $16\rightarrow 8\rightarrow 4$) into smaller blocks in the subsequent step.

For each region $R_{i,j}$, we compute its privacy sensitivity metric - the region-variance $RV_{i,j}^2$ - through:
\vspace{-0.08in}
\begin{equation}
RV_{i,j}^2 = \frac{1}{C}\sum_{c\in C} \left[ \frac{1}{|R_{i,j}|} \sum_{(x,y)\in R_{i,j}} (I_{x,y,c} - \mu_{i,j,c})^2 \right]
\label{eq:region_variance}
\end{equation}
\noindent where:
\begin{align*}
% C \quad &\colon \text{Number of Channels in the image}\\
I_{x,y,c} &\colon \text{Pixel value at position }(x,y)\text{ in channel }c \\
\mu_{i,j,c} &= \frac{1}{|R_{i,j}|}\sum_{(x,y)\in R_{i,j}} I_{x,y,c} \quad \text{(Mean value in channel c)} \\
|R_{i,j}| &= S^2 \quad \text{(Number of pixels per region)}
\end{align*}

\textbf{2) Variance-Guided Adaptive Region Partitioning.}
Building upon the region-variance calculations, CSVAR implements an adaptive partitioning strategy that automatically adjusts protection strength based on each region's privacy sensitivity. Specifically, CSVAR will partition a privacy-sensitive region with high variance into small blocks to enhance privacy protection. Meanwhile, for a privacy-insensitive region with low variance, CSVAR will partition it with much bigger block sizes to preserve model utility.
% This variance-guided analysis enables adaptive processing: high-variance regions (indicating privacy-sensitive content like facial features) are divided into smaller blocks for stronger protection, while low-variance regions (typically background or homogeneous areas) use larger blocks to preserve structural integrity.

CSVAR first computes the median variance $\tilde{RV}^2$ across all regions to establish the sensitivity threshold. For each region $R_{i,j}$:

\begin{equation}
\text{BlockSize} = 
\begin{cases}
\lfloor S/4 \rfloor  & \text{if } RV_{i,j}^2 > \tilde{RV}^2 \text{ (privacy-sensitive)} \\
\lfloor S/2 \rfloor  & \text{if } RV_{i,j}^2 \leq \tilde{RV}^2 \text{ (privacy-insensitive)}
\end{cases}
\label{eq:block_partition}
\end{equation}

\textbf{3) Channel-wise Spatial Shuffling between Blocks.}
After adaptive block partitioning, CSVAR performs two complementary shuffling operations to protect visual privacy. First, \textit{spatial shuffling} randomly permutes the positions of all blocks within each region using different random seeds for each training epoch. This breaks spatial correlations between blocks while preserving pixel relationships within each block. 

Second, \textit{channel-wise shuffling} processes all channels independently: each block is decomposed into per-channel sub-blocks (e.g., R/G/B for color images), which are then shuffled across the region. For example, a block's first channel may relocate to the region's top-left while other channels scatter to different positions. This dispersion breaks channel correlations to disrupt color distributions, as adjacent positions now contain uncorrelated channels from different blocks.

\section{evaluation}
\begin{figure}
    \centering %图片居中
    \includegraphics[width=0.48\textwidth]{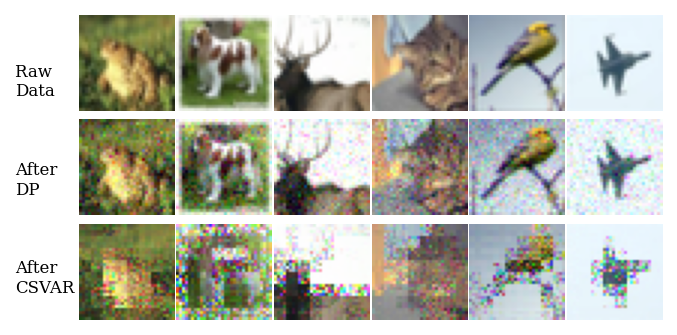}
    %插入图片，[]中设置图片大小，{}中是图片文件名
    \vspace{-0.2in}
    \caption{Comparison of Visual obfuscation effect with different Protections.} %最终文档中希望显示的图片标题
    \label{img-effect} %用于文内引用的标签
    \vspace{-0.2in}
\end{figure}
In this section, we conduct extensive experiments across various real-world datasets and models to evaluate CSVAR's effectiveness in enhancing visual privacy and preservation in model utility.  We begin with the experimental setups, then present evaluation results to answer the following questions:
% including the models, datasets, and baselines used in our experiments. Subsequently, we present the result of our evaluation experiments to answer the following questions:
\begin{itemize}
    \item Can CSVAR enhance visual privacy to generate visually obfuscated images that exhibit high perceptual ambiguity to human eyes and mitigate the effectiveness of adversarial data reconstruction attacks?
    \item Can CSVAR achieve a good trade-off
        between visual privacy protection and model utility?
\end{itemize}
\subsection{Experiment Setup}
\textbf{Models.}
We evaluate CSVAR using three representative CNN architectures selected to span the spectrum of modern computer vision applications. ResNet-50 serves as the base residual network, representing standard medium-scale models widely used in FL systems. MobileNet is included as the lightweight architecture optimized for edge devices with limited computational resources. ShuffleNet provides an additional efficiency-focused design point. This selection covers a range of model capacities (4.2M to 25.5M parameters) and computational requirements (0.6B to 4.1B FLOPs), ensuring thorough evaluation of CSVAR's compatibility across different neural network designs commonly deployed in computer vision applications under federated learning scenarios.

% Table \cite{tab} presents the datasets used in our work. Four representative datasets in the CV domain were selected for testing. Specifically, ImageNet-100 is a standard public dataset released by the well-known Kaggle contest [1]. And CIFAR-10 is a color image dataset consisting of 10 categories, with a total of 60,000 images; MNIST is a dataset for handwritten digit recognition, containing 60,000 training samples and 10,000 testing samples; CelebA dataset, also known as the ORL (Olivetti Research Lab) face database, consists of 400 grayscale face images belonging to 40 different individuals, with 10 images per individual. This dataset is commonly used for face recognition and face detection tasks. The selection criteria for the DNN models used in this work are as follows: (1) the test models should encompass a variety of mainstream frameworks for CV tasks, including Transformer structures (e.g., ViT-B [9] and Swin-T [35]), directly connected CNNs (e.g., AlexNet [29] and VGG [44]), and residual network models (e.g., ResNet [18] and DenseNet [23]); (2) the DNN models should cover network models with varying parameter scales and computational efforts, including very large networks like ViT-B, large networks like VGG, and lightweight networks like MobileNet [22] and ShuffleNet [52].
\textbf{Datasets.}
Our experiments employ three benchmark datasets widely used in computer vision. CelebA Face Dataset (400 grayscale images of 40 subjects) evaluates the effectiveness of facial privacy protection. MNIST (70,000 handwritten digits) can assess preservation of basic structural features while preventing digit recognition. CIFAR-10 (60,000 color images across 10 categories) can test performance on more complex natural images with varied textures and compositions. %This combination provides comprehensive coverage of image types (grayscale to color), and feature complexities (simple edges to detailed textures).

\textbf{Baselines.} We compare CSVAR against two baselines: (1) \textit{Vanilla FL}, the standard federated learning framework without any privacy protection, serving as the upper-bound reference for model utility; and (2) \textit{DP-enhanced FL}, which adds Gaussian noise to training images with a Differential-Privacy based method. Here the $\sigma$ of Gaussian noise is 50.
\subsection{Enhancement for Visual Privacy}
\begin{figure}
    \centering %图片居中
    \includegraphics[width=0.4\textwidth]{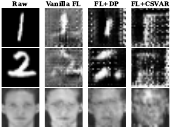}
    %插入图片，[]中设置图片大小，{}中是图片文件名
    \vspace{-0.05in}
    \caption{Effect of GAN-based Data Reconstruction attack.} %最终文档中希望显示的图片标题
    \label{fig:gan} %用于文内引用的标签
    \vspace{-0.15in}
\end{figure}
\textbf{Visual Obfuscation Effectiveness.}
Figure \ref{img-effect} demonstrates the visual obfuscation effects on CIFAR-10 images across three approaches. Vanilla FL uses raw training images with no obfuscation, leaving all object details clearly visible. FL+DP($\sigma=50$) applies random noise to all pixels, yet fails to adequately obscure privacy-sensitive regions - object shapes like airplanes and animals remain distinguishable. In contrast, CSVAR's adaptive shuffling obfuscates images to show complete disruption of original object contours and textures, with no identifiable features remaining. This figure confirms CSVAR's superior perceptual ambiguity, particularly in preserving privacy for sensitive regions that DP fails to obfuscate adequately.

\textbf{Resistance to GAN-based Reconstruction attacks. }  
We evaluate GAN-based data reconstruction attacks \cite{hitaj2017deep} on models trained with different protection schemes. As shown in Figure \ref{fig:gan}, attacking models trained by vanilla FL yields nearly perfect reconstructions where all facial features from CelebA and number features from MNIST are clearly recoverable. 
While FL+DP-protected models produce visibly distorted outputs, salient facial and number features still remain decipherable to human eyes.
With FL+CSVAR, the reconstructed images show only random noise patterns - no eyes, noses can be distinguished in facial images. 
% We evaluate GAN-based data reconstruction attacks \cite{DBLP:conf/ccs/HitajAP17} on models trained with different protection schemes. As shown in Figure \ref{fig:gan}, attacking models trained by vanilla FL yields nearly perfect reconstructions where number features from MNIST are clearly recoverable. 
% While FL+DP-protected models produce visibly distorted outputs, critical semantic components, particularly salient number features, remain decipherable to human observers.
% With the integration of FL+CSVAR, the reconstructed images exhibit merely random noise patterns, rendering it impossible to distinguish any numerical components within the recovered images.

\begin{figure}
    \centering %图片居中
    \includegraphics[width=0.47\textwidth]{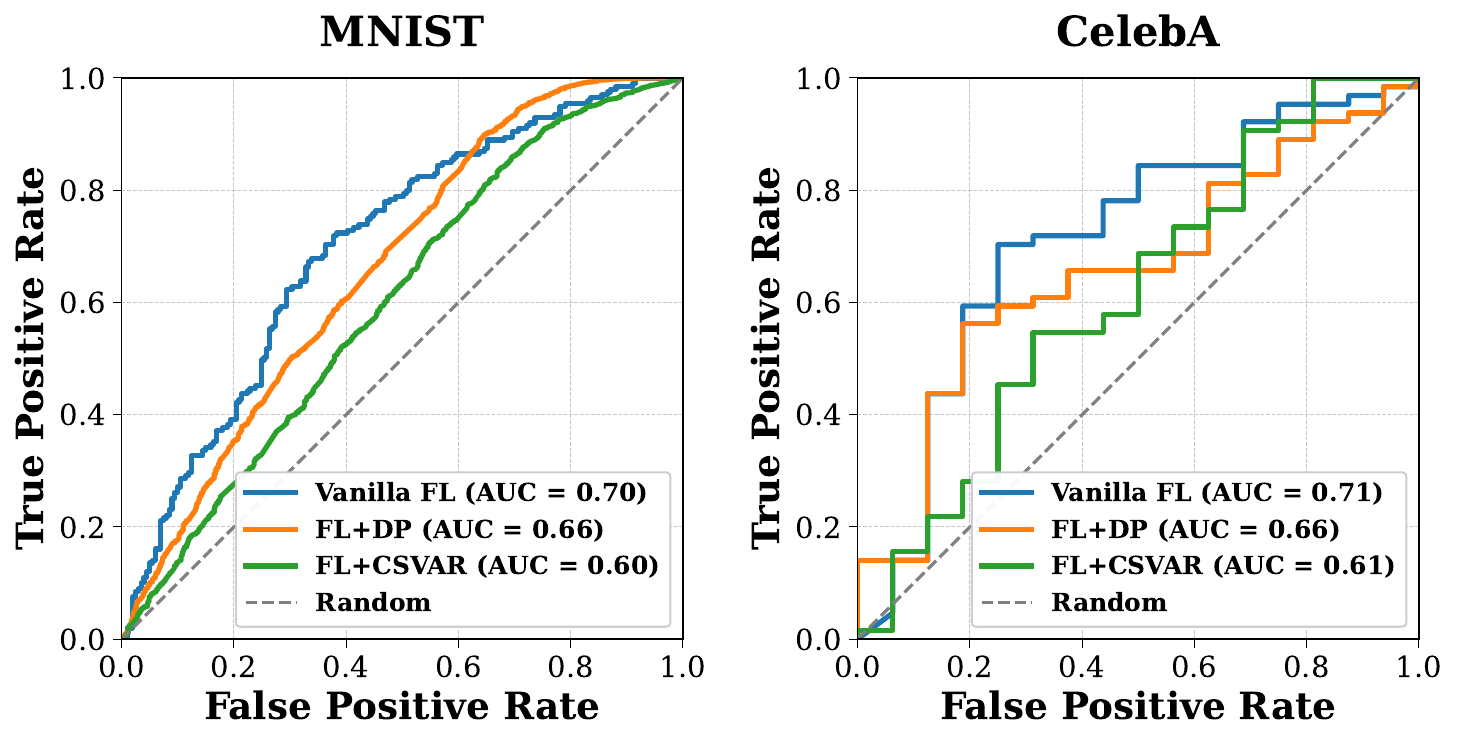}
    %插入图片，[]中设置图片大小，{}中是图片文件名
    \vspace{-0.05in}
    \caption{ROC results of MIA with different protection methods. Low AUC means low attack Success Rate.} %最终文档中希望显示的图片标题
    \label{img-eval-mia} %用于文内引用的标签
    \vspace{-0.2in}
\end{figure}
\textbf{Mitigation of Membership inference attacks.} 
Targeting the model weight after training, membership Inference attacks aim to determine whether a given data point belongs to the training dataset. 
Figure \ref{img-eval-mia} demonstrates that membership inference attacks achieve high accuracy when targeting models trained with vanilla FL(the blue curve, AUC=0.7). For FL+DP, the attack performance decreases significantly(the orange curve, AUC=0.66), though some leakage remains detectable. CSVAR provides the strongest protection, where the attack AUC decreases to 0.6, close to the random guess curve.
\subsection{Model Utility Preservation}
We evaluate CSVAR's impact on model accuracy across three standard datasets (MNIST, CelebA, CIFAR-10) and three models (ResNet50, MobileNet, ShuffleNet), comparing against vanilla FL and FL+DP baselines.
The results demonstrate CSVAR's ability to maintain model utility while providing strong visual privacy protection.

\textbf{Accuracy Analysis.}
As demonstrated in Table \ref{tab}, CSVAR maintains model performance close to models trained with vanilla FL despite its strong privacy protections, in stark contrast to the substantial accuracy degradation with FL+DP. In general, on MNIST and CIFAR10 datasets, models trained with FL-CSVAR introduce negligible model utility loss(average 0.21\%) compared to vanilla FL.
Furthermore, across all evaluated models, FL+CSVAR demonstrates a consistent accuracy improvement over FL+DP, with an average accuracy increase of 3.23\% on the CIFAR10 and a more pronounced 9.75\% increase on the CelebA face dataset.  
Most notably, on CelebA, FL+CSVAR achieves 85.25\% accuracy with ResNet50—demonstrating a mere 4\% decrease from vanilla FL, while FL+DP exhibits a catastrophic 24.25\% accuracy drop. This stark contrast underscores CSVAR's exceptional capability to preserve model utility while enhancing privacy.

\begin{table}[]
\centering
\begin{tabular}{l|lrrr}
\hline
        &            & \multicolumn{1}{l}{Resnet-50} & \multicolumn{1}{l}{ShuffleNet} & \multicolumn{1}{l}{MobileNet} \\ \hline
        & Vanilla FL & 97.42\%                       & 96.88\%                        & 97.08\%                       \\
MNIST   & FL+DP      & 97.09\%                       & 95.44\%                        & 96.92\%                       \\
        & FL+CSVAR   & 97.39\%                       & 95.67\%                        & 97.03\%                       \\ \hline
        & Vanilla FL & 84.24\%                       & 85.09\%                        & 85.18\%                       \\
CIFAR10 & FL+DP      & 80.49\%                       & 82.33\%                        & 80.03\%                       \\
        & FL+CSVAR   & 84.17\%                       & 83.71\%                        & 84.67\%                       \\ \hline
        & Vanilla FL & 89.75\%                       & 85.50\%                        & 86.75\%                       \\
CelebA   & FL+DP      & 65.50\%                       & 77.25\%                        & 78.75\%                       \\
        & FL+CSVAR   & 85.25\%                       & 82.74\%                        & 82.75\%                       \\ \hline
\end{tabular}
\vspace{0.03in}
\caption{Model Accuracy trained with different Protection Methods.}
\label{tab}
\vspace{-0.2in}
\end{table}

\section{conclusion}
\addtolength{\topmargin}{0.05in}
We propose CSVAR, a novel image shuffling framework
to prevent visual privacy leakage from overfitting and the
Client-side insecure private data transmission under Federated Learning. CSVAR adopts region-variance as the metric to measure a region's visual privacy sensitivity.  CSVAR adaptively partitions each image region into smaller blocks with different granularity guided by region-variance, and then shuffles between blocks in both the spatial domains and chromatic channels. Experimental results show that CSVAR achieves a good trade-off between visual privacy protection against overfitting-induced privacy leakage and model utility.

\bibliographystyle{IEEEtran}  % 调用BST文件
\bibliography{src/reference}    % 指定.bib文件名（无需后缀）

\vspace{12pt}
% \color{red}
% IEEE conference templates contain guidance text for composing and formatting conference papers. Please ensure that all template text is removed from your conference paper prior to submission to the conference. Failure to remove the template text from your paper may result in your paper not being published.

\end{document}